\documentclass[%
 reprint,
 onecolumn,
superscriptaddress,
 amsmath,amssymb,
 aps,
 prx,
]{revtex4-2}
\usepackage{physics}

\makeatletter
\let\ORIbbl@fixname\bbl@fixname
\def\bbl@fixname#1{%
  \@ifundefined{languagealias@\expandafter\string#1}
    {\ORIbbl@fixname#1}
    {\edef\languagename{\@nameuse{languagealias@#1}}}%
}
\newcommand{\definelanguagealias}[2]{%
  \@namedef{languagealias@#1}{#2}%
}
\makeatother

\usepackage[letterpaper,top=2cm,bottom=2cm,left=3cm,right=3cm,marginparwidth=1.75cm]{geometry}

\usepackage{amsmath}
\usepackage{amssymb}
\usepackage{amsthm}
\usepackage{bm}
\usepackage{graphicx}
\usepackage[colorlinks=true, allcolors=blue]{hyperref}

\usepackage[caption=false]{subfig}
\usepackage{algorithm2e}
\RestyleAlgo{ruled}

\newcommand{\btheta}{{\bm{\theta}}}
\newcommand{\bx}{{\bm{x}}}

\newcommand{\bphi}{\bm{\phi}}
\newcommand{\bPhi}{\bm{\Phi}}
\newcommand{\bTheta}{\bm{\Theta}}

\newcommand{\SU}[1]{\mathrm{SU}(#1)}
\newcommand{\su}[1]{\mathfrak{su}(#1)}
\DeclareMathOperator*{\argmin}{\arg\!\min}

\begin{document}
    
\title{Pulse Quality Optimisation in Quantum Optimal Control}

\author{Dylan Lewis}
\email{dlewis3@imperial.ac.uk}
\affiliation{Blackett Laboratory, Imperial College London, London SW7 2AZ, U.K.}

\author{Roeland Wiersema}
\email{rwiersema@flatironinstitute.org}
\affiliation{Center for Computational Quantum Physics, Flatiron Institute, 162 Fifth Avenue, New York, NY 10010, USA}

\begin{abstract}
Quantum optimal control methods are widely used to design experimental control pulses such as laser amplitudes, phases, or detunings, that implement a target unitary evolution. In practice, what makes a pulse “good” depends not only on its fidelity, but also on the experimental setting and the relevant hardware constraints. Here, we introduce geometric quantum control with kernel optimisation (GECKO), a model-agnostic method for improving control pulses after a high-fidelity solution has been found. GECKO uses the Riemannian geometry of the special unitary group to identify directions in pulse space that leave the implemented unitary unchanged to first order, allowing one to traverse level sets of the control landscape while optimising a chosen differentiable pulse-quality function. We demonstrate GECKO on a transverse-field Ising Hamiltonian implementing CZ and CNOT gates, optimising pulse properties including spectral filtering, smoothness, robustness to parameter deviations, and pulse duration. In all cases, GECKO finds substantially improved pulse solutions.
\end{abstract}

\maketitle

\section{\label{sec:intro}Introduction} 
Controlling quantum systems to implement desired transformations is an essential component of emerging quantum technologies~\cite{kochQuantumOptimalControl2022}. Achieving such control requires precisely shaped pulses that steer the system dynamics towards a desired target evolution. Control protocols for full unitary dynamics, as well as simpler state-preparation tasks, have been designed for a wide range of systems and are now routinely used in quantum hardware~\cite{brif_control_2010,kelly_optimal_2014, krantz_quantum_2019, bruzewicz_trapped-ion_2019, evered_high-fidelity_2023}. However, each hardware platform imposes distinct constraints on the available controls, pulse amplitudes, bandwidths, and parametrisations, requiring new pulses to be designed for each experimental setup.

Quantum optimal control theory is a set of tools for designing control pulses that realize a desired unitary or state transformation. Standard methods are typically based on gradient ascent of a fidelity function, notably gradient ascent pulse engineering (GRAPE)~\cite{khanejaOptimalControlCoupled2005}, Krotov's method~\cite{krotovGlobalMethodsOptimal1993, palaoOptimalControlTheory2003,reich2012monotonically, morzhinKrotovMethodOptimal2019} and the chopped random basis (CRAB) approach~\cite{canevaChoppedRandombasisQuantum2011, mullerOneDecadeQuantum2022}. Recently, we introduced  geodesic pulse engineering (GEOPE), which exploits geodesics on the Riemannian manifold $\SU{2^n}$ to guide iterative control-parameter updates for $n$-qubit systems~\cite{lewisGeodesicAlgorithmUnitary2025, lewis2025quantum}. 
While unconstrained quantum control landscapes are often favourable under suitable controllability assumptions, the introduction of practical constraints can qualitatively change the optimisation problem. Amplitude, bandwidth, and parametrisation restrictions may introduce artificial traps, reduce the set of accessible controls, or slow convergence substantially, even when high-fidelity solutions exist~\cite{pechen2011there, pechen2012quantum, pechen2012trap, zhdanov2015role}. At the same time, quantum control problems typically admit many distinct controls that realise the same target with comparable fidelity~\cite{chakrabarti2007quantum, roslund2006laboratory,larocca2018quantum}.

These observations suggest an alternative strategy for addressing the constrained control problem: rather than imposing all experimental requirements during the search for a control solution, one may first find a high-fidelity pulse and then move along the corresponding level set to improve secondary pulse properties. Earlier works have shown that such level-set exploration is possible through observable-preserving flows~\cite{rothman2005observable}, D-MORPH methods~\cite{beltrani2011exploring,fan2025top}, and other landscape-geometric constructions~\cite{rothman2006exploring, larocca2020exploiting}, revealing connected families of controls with identical or near-identical performance. Nevertheless, these approaches are often problem-specific and are not naturally formulated for general pulse-quality objectives in gate-design settings. 

In this work, we introduce geometric quantum control with kernel optimisation (GECKO), a model-agnostic method for optimising secondary pulse-quality objectives along level sets of the quantum control landscape.
Starting from a control pulse that generates the target unitary transformation, GECKO performs constrained optimisation along the corresponding level set of the quantum control landscape. This is achieved by identifying the nullspace directions tangent to the fidelity level set, which correspond to pulse updates that leave the implemented unitary unchanged to first order. We illustrate the power of GECKO with several numerical examples, including spectral filtering to remove high-frequency components, pulse smoothing, improved robustness to experimental noise, and the reduction of the duration of a pulse toward a time-optimal solution.

\section{\label{sec:kernel_optimisation}Geometric quantum control with Kernel optimisation}

\subsection{Pulse quality optimisation with constant fidelity}

Consider a target unitary $U_{\mathrm{target}}$ acting on $n$ qubits which we seek to approximate using a sequence of $L$ piecewise-constant pulses,
\begin{align}
    U_\textrm{G}(\bPhi) = U(\bphi_L) U(\bphi_{L-1}) \cdots U(\bphi_2) U(\bphi_1).
\end{align}
Each pulse segment is generated by a Hamiltonian with $K$ controllable terms,
\begin{align}
    H_l(\bphi_l) = \left( \sum_{d=1}^D g_{l,d}  F_d + \sum_{k=1}^K \phi_{l,k}  G_k \right),
\end{align}
where $G_k\in\mathcal{H}_\textrm{C}\subseteq\mathcal{P}_n \setminus\{I\}$ are control generators and $\mathcal{P}_n$ denotes the set of Pauli operators acting on $n$ qubits.
In addition, the pulse can contain $D$ drift terms $F_d \in \mathcal{H}_\textrm{D} \subseteq\mathcal{P}_n \setminus\{I\}$ determined by the experimental setup. The Hamiltonian $H_l(\bphi_l)$ then implements the unitary
\begin{align*}
     U(\bphi_l) = \exp\left\{-\frac{i}{\hbar} H_l(\bphi_l)\Delta t \right\}.
\end{align*}
Each pulse segment has duration $\Delta t$, so the total pulse duration is $T=L\Delta t$.
We set the Planck constant $\hbar=1$, so that $g_{l,d}$ and $\phi_{l,k}$ have units of frequency. 

We consider a set of parameters $\bPhi \in \mathbb{R}^{LK}$ to be a solution if the fidelity is 1 within a small predetermined error $\varepsilon$,
\begin{align}\label{eq:fid}
    F(\bPhi,U_{\mathrm{target}}) = \frac{1}{N}\big|\mathrm{Tr}\{U_{\mathrm G}^{\dagger}(\bPhi)U_{\mathrm{target}}\}\big| > 1 - \varepsilon,
\end{align}
where $N=2^n$ is the Hilbert-space dimension.
A solution can be found using a variety of quantum optimal control methods~\cite{khanejaOptimalControlCoupled2005,krotovGlobalMethodsOptimal1993, palaoOptimalControlTheory2003,reich2012monotonically, morzhinKrotovMethodOptimal2019,canevaChoppedRandombasisQuantum2011, mullerOneDecadeQuantum2022,lewis2025quantum}. However, the solutions $\bPhi$ found with these methods often do not satisfy hardware constraints, which can vary greatly between specific experimental setups. As a result, high-fidelity pulses may still be experimentally difficult to implement.

We therefore seek alternative high-fidelity pulse solutions that also optimise a secondary objective $\mathcal{Q}(\bPhi)$ that quantifies the quality of the pulse. 
To this end, we introduce geometric quantum control with kernel optimisation (GECKO):  a constrained optimisation method where we attempt to minimise the value of a quality function $\mathcal{Q}(\bPhi)$ while maintaining high fidelity with the target unitary:
\begin{align}\label{eq:Qmin}
     \min_{\boldsymbol{\Phi}} ~  \mathcal{Q}(\boldsymbol{\Phi}) \qquad
    \text{subject to} \qquad  F(\boldsymbol{\Phi}, U_{\mathrm{target}}) > 1 - \varepsilon.
\end{align}
We assume only that the quality function $\mathcal{Q}(\bPhi):\mathbb{R}^{LK}\to\mathbb{R}$ is smooth and that we can obtain its gradient $\nabla \mathcal{Q}(\bPhi)$ via automatic differentiation. Constrained optimisation problems of this form can be addressed by many methods; see Ref.~\cite{venter2010} for an overview. Here, we pursue a geometric approach that exploits the large number of unused degrees of freedom in the pulse parametrisation.

\subsection{A Kernel method to traverse the fidelity level sets}
To decrease $\mathcal{Q}(\bPhi)$ while preserving fidelity, we must identify parameter updates that remain tangent to the level set of $F(\bPhi,U_{\mathrm{target}})$.
We can formalise this notion by considering perturbations $\bm{\delta\Phi}\in T_{\bPhi} \mathbb{R}^{LK}$ of the parameters $\bPhi$ and determining when 
\begin{align}\label{eq:constraint}
    F(\boldsymbol{\Phi}, U_{\mathrm{target}})=F(\boldsymbol{\Phi} + \bm{\delta\Phi}, U_{\mathrm{target}}) ,\qquad \bm{\delta\Phi}\neq0.
\end{align}
A sufficient, stronger condition for Eq.~\eqref{eq:constraint} to hold is that the implemented unitary remain unchanged up to a global phase,
\begin{align}
    U_{\mathrm{G}}(\bPhi) = e^{i\varphi} U_{\mathrm{G}}(\bPhi + \bm{\delta\Phi}) ,\quad \varphi\in\mathbb{R}.
\end{align}
The Taylor expansion of $U_{\mathrm{G}}(\bPhi + \bm{\delta}\bPhi)$ for small $\Vert\bm{\delta\Phi} \Vert$ about $\bPhi$ is given by
\begin{align}
    U_{\mathrm{G}}(\bPhi + \bm{\delta}\bPhi) = U_{\mathrm{G}}(\bPhi) + \sum_{l=1}^{L}\sum_{k=1}^{K}\delta\phi_{l,k} \frac{\partial U_{\mathrm{G}}(\bPhi)}{\partial \phi_{l,k}} + O(\Vert\bm{\delta\Phi}\Vert^2).
\end{align}
Hence the requirement that $U_{\mathrm{G}}(\bPhi) = e^{i\varphi} U_{\mathrm{G}}(\bPhi + \bm{\delta\Phi})$ up to first order directly implies that
\begin{align}\label{eq:null_space}
    \sum_{l=1}^{L}\sum_{k=1}^{K}\mathbf{J}_{l,k}(\bPhi)\delta\phi_{l,k}  = 0,\qquad
    \mathbf{J}_{l,k}(\bPhi) = \frac{\partial U_\textrm{G}(\bPhi)}{\partial \phi_{l,k}} \in T_{U_\textrm{G}(\bPhi)} \SU{N}.
\end{align}
The map $\mathbf{J}(\bPhi): T_{\bPhi}\mathbb{R}^{LK}\to T_{U_\textrm{G}(\bPhi)} \SU{N}$ is a linear operator between the tangent space of the pulse parameters and the tangent space of the group $\SU{N}$. Both of these spaces can be identified with corresponding real vector spaces $V$ and $W$ by choosing a basis, $T_{\bPhi}\mathbb{R}^{LK}\cong \mathbb{R}^{LK}$ and $T_{U_\textrm{G}(\bPhi)} \SU{N} \cong \su{N}\cong \mathbb{R}^{N^2-1}$.
As a linear map, $\mathbf{J}(\bPhi)$ has a kernel, or nullspace, defined by 
\begin{align}
    \ker(\mathbf{J}(\bPhi))=\{\bx\in V: \mathbf{J}(\bPhi)\bx = 0\}.
\end{align}
Note that this is exactly the requirement of Eq.~\eqref{eq:null_space}: the directions $\bm{\delta\Phi}\in \ker(\mathbf{J}(\bPhi))$ keep $U_\textrm{G}(\boldsymbol{\Phi})$ unchanged up to first order, which results in constant fidelity up to first order.

\begin{figure}[htb!]
    \centering
    \includegraphics[width=\linewidth]{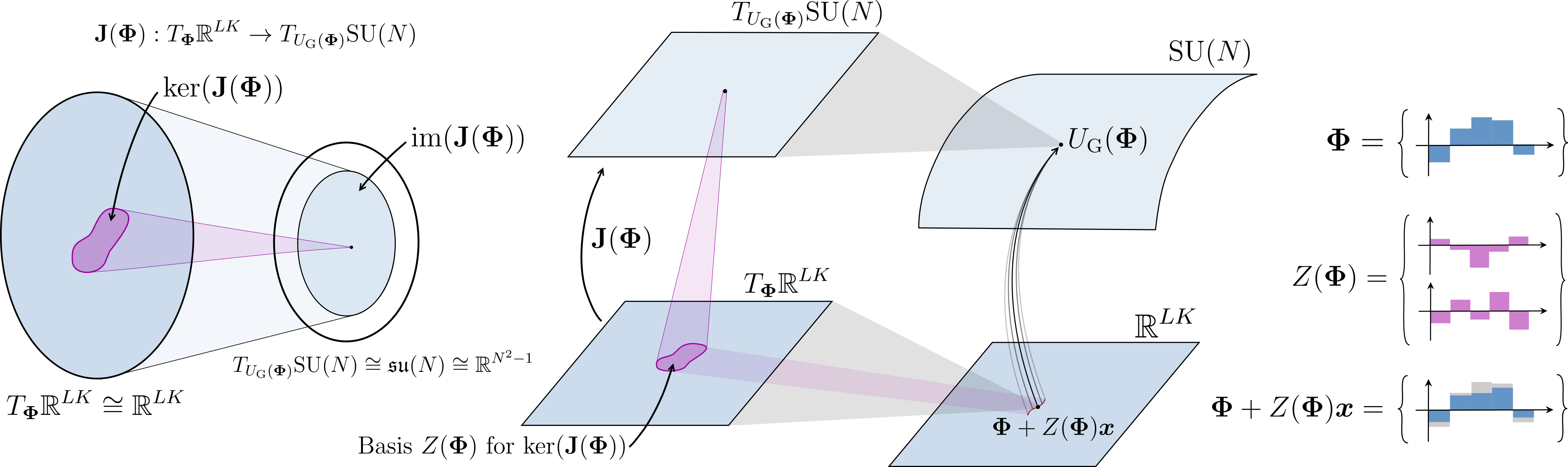}
    \caption{The unitary $U_\mathrm{G}(\bPhi)$ is an element of the special unitary group $\SU{N}$, and a point on a smooth manifold. $U_\mathrm{G}(\bPhi)$ is also a map from parameters $\bPhi$, in the space $\mathbb{R}^{LK}$, to the special unitary group. The Jacobian $\mathbf{J}(\bPhi)$ is a map from the tangent space of the parameter space $T_{\bPhi} \mathbb{R}^{LK}$ to the tangent space of the special unitary group $T_{U_\mathrm{G}(\bPhi)} \SU{N}$. The kernel of the Jacobian $\mathbf{J}(\bPhi)$ is then spanned by the tangent vectors in $T_{\bPhi} \mathbb{R}^{LK}$ that get mapped to the zero vector in $T_{U_\mathrm{G}(\bPhi)} \SU{N}$. The leftmost illustration shows how, if $\dim(T_{\bPhi}\mathbb{R}^{LK}) > \dim(T_{U_{\textrm{G}}(\bPhi)}\SU{N})$, the Jacobian map has a non-trivial kernel. The kernel can be described by an orthonormal set of basis vectors, $Z(\bPhi)$. This illustrates the key idea in GECKO, which is to find a set of vectors $\bm{\delta\Phi} = Z(\bPhi)\bx$ that can update $\bPhi$ such that, up to first order, the element in the special unitary group remains unchanged, i.e. $U_\mathrm{G}(\bPhi) = U_\mathrm{G}(\bPhi+\bm{\delta\Phi}) + O(\Vert \bm{\delta\Phi}\Vert^2)$. The coordinates $\bm{x}$ refer to the coefficients of the parameter tangent space kernel vectors that keep the element of the special unitary invariant up to first order. The vector $\bm{x}$ is found such that a specific cost function of the pulse, $\mathcal{Q}(\bPhi)$, is optimised. The plots, on the right, are illustrations of a simple example for a parameter vector $\bPhi$ (with $K=1$). Where a choice of $\bm{x}$, with sufficiently small magnitude, changes the parameter vector to either minimise or maximise $\mathcal{Q}(\bPhi)$, but does not change the point on the special unitary manifold.}
    \label{fig:maths_diagram}
\end{figure}

By the rank-nullity theorem, the kernel $\ker(\mathbf{J}(\bPhi))\subset V$ is a vector subspace of $V$ with dimension
\begin{align}
    R = \dim(\ker(\mathbf{J}(\bPhi))) = LK - \rank(\mathbf{J}(\bPhi)),
\end{align}
giving $R\geq LK-N^2+1$, since the rank of $\mathbf{J}(\bPhi)$ is at most $\dim(T_{U_\textrm{G}(\bPhi)} \SU{N}) = N^2-1$. We can define an orthonormal basis $Z(\bPhi)\in\mathbb{R}^{LK\times R}$ for the kernel such that $Z(\bPhi)^T Z(\bPhi)=I$. Thus, every first-order, fidelity-preserving update can then be written as
\begin{align}\label{eq:Zx}
    \bm{\delta\Phi} = Z(\bPhi) \bx \in \ker(\mathbf{J}(\bPhi)), \quad \forall \bx\in\mathbb{R}^{R},
\end{align}
giving us a way to directly parameterise the kernel directions with the parameters $\bx$. Fig.~\ref{fig:maths_diagram} illustrates this construction.
Related Jacobian nullspace approaches to traversing solution manifolds have been used in robotics~\cite{khatib2003unified,dietrich2015overview}, shape optimisation~\cite{feppon2020null, feppon2024density}, and protein dynamics~\cite{pachov2015nullspace}.

\subsection{Solving the constrained pulse-quality optimisation problem}
We now have a way to directly parameterise updates to the solution $\bPhi$ that preserve $U_\textrm{G}(\bPhi)$ up to first order using Eq.~\eqref{eq:Zx}. This kernel parameterisation leads to two natural update strategies for the optimisation problem of Eq.~\eqref{eq:Qmin}.
In the first approach, we fix the kernel basis $Z(\bPhi)$ at the current solution and optimise $\mathcal{Q}$ over the corresponding local kernel coordinates:
\begin{align}\label{eq:Qminx}
     \min_{\boldsymbol{\bx}} \quad \mathcal{Q}( \bPhi + Z(\bPhi) \bx).
\end{align}
Depending on the form of $\mathcal{Q}$, this reduced problem can be solved either directly, as in Section~\ref{sec:smoothing}, or by gradient descent in the kernel coordinates. In the latter case, we compute $\nabla_{\bx}\mathcal{Q}(\bPhi+Z(\bPhi)\bx)$ and update $\bx$ in a descent direction:
\begin{align*}
    \boldsymbol{\Delta} \bx  = -\nabla_\bx \mathcal{Q}( \bPhi + Z(\bPhi) \bx).
\end{align*}
For the second approach, we can work in the original formulation of the optimisation problem, given in Eq.~\eqref{eq:Qmin}, and project the gradient $\nabla \mathcal{Q}(\bPhi)$ onto the set of feasible directions~\cite{luenberger1972gradient,shikhman2009constrained, v2025gradient}. 
Here, the feasible first-order directions are precisely the kernel directions parameterised by Eq.~\eqref{eq:Zx}. To achieve this, we find the kernel vector closest to the negative gradient $- \nabla \mathcal{Q}(\bPhi)$ and solve the subproblem
\begin{align}\label{eq:optx}
     \boldsymbol{\Delta} \bx = \argmin_{\bx} \quad & \Vert Z(\bPhi) \bx + \nabla \mathcal{Q}(\bPhi)\Vert^2,
\end{align}
which is a linear least-squares problem with solution $\boldsymbol{\Delta} \bx = -Z(\bPhi)^T \nabla \mathcal{Q}(\bPhi)$~\cite{boyd2004convex}.

Once we have found a direction $\boldsymbol{\Delta} \bx$ with either approach, we form the corresponding parameter-space update and normalise it to a prescribed step size $s$:
\begin{align}
    \bPhi \mapsto \bPhi + \bm{\Delta\bPhi} ,\qquad \bm{\Delta\bPhi} = s\frac{Z(\bPhi) \boldsymbol{\Delta} \bx}{\norm{Z(\bPhi) \boldsymbol{\Delta} \bx}}.
    \label{eq:update_step}
\end{align}
Because the update lies in the kernel of the Jacobian, the induced change in the implemented unitary vanishes to first order; hence the deviation in $U_\textrm{G}$ after the update is $O(\Vert\Delta\bPhi\Vert^2)$.
Note that this error can be controlled by an appropriate choice of $s$. Larger values of $s$ may reduce $\mathcal{Q}$ more rapidly, but also produce larger second-order deviations from the fidelity level set. If, as a result of these errors, we violate the constraint $F(\bPhi^\prime, U_\mathrm{target})>1-\varepsilon$, we  re-optimise the parameters with our chosen quantum optimal control method. The resulting algorithm iterates between GECKO updates and fidelity-restoring control optimisation until a high-fidelity pulse with sufficiently low $\mathcal{Q}$ is obtained. The general algorithm for GECKO is given in Algorithm~\ref{alg:geometric_pulse_smoothing}.

The cost of one GECKO step has several contributions. Computing the pulse unitary costs $O(LN^3)$. Computing the Jacobian and projecting it onto the chosen real vector-space representation costs $O(LKN^4)$. Extracting the kernel via a singular value decomposition costs $O(LKN^4)$ in the scaling considered here. The cost of computing $\nabla\mathcal{Q}$ depends on the chosen quality function: it can be as large as $O(LKN^3)$ if unitary derivatives are required, but is only $O(LK)$ for local pulse-space objectives and $O(LK\log L)$ for Fourier-domain objectives. Finally, solving the least-squares problem costs $O(LKR^2)$ in the worst case~\cite{boyd2004convex}. After each update, an additional fidelity calculation is required to check whether the updated pulse still satisfies the constraint, which costs $O(LN^3)$.
\begin{algorithm}
\caption{Geometric quantum control with kernel optimisation (GECKO) algorithm.}\label{alg:geometric_pulse_smoothing}
\KwIn{$\bm{\bPhi}$, $\mathcal{H}$, $\mathcal{Q}$, $s$,  $L$, $n_S$, $\mathcal{Q}_\textrm{aim}$}
\KwOut{$\bm{\Phi}$}
$K \gets |\mathcal{H}|$ \\
\text{Obtain the Jacobian function:}\\
\For{$l \in (1,\ldots, L$)}{
\For{$k \in (1,\ldots, K)$}{
$dU_{\mathrm{G},l,k}(\bTheta) = \partial_{\theta_{l,k}} \mathfrak{Re}[U(\btheta_L)\cdots U(\btheta_1)] + i \partial_{\theta_{l,k}}\mathfrak{Im}[U(\btheta_L)\cdots U(\btheta_1)]$}}

\text{Current evolution is given by:} $U_{\mathrm{G}}(\bPhi) = U(\bphi_L) U(\bphi_{L-1})\cdots U(\bphi_1)$ \\
\text{Define pulse quality:} $\mathcal{Q}(\bPhi)$

\For{$m \in \{ 1, 2, ..., n_S\}$}{
    \text{Evaluate the Jacobian at $\bPhi$:}\\

    {~~$\mathbf{J}_{l,k}(\bPhi) \gets dU_{\mathrm{G},l,k}(\bTheta)|_{\bPhi}$
    
    ~~$\bm{j}_{l,k} \gets \sum_{j=1}^{N^2 -1} \Tr{G_j \mathbf{J}_{l,k}(\bPhi)} \hat{e}_j$} \\
    ~~$J \gets \textrm{flatten}~\bm{j}_{l,k}~\textrm{along}~l,k$
    
    \text{Find the kernel basis vectors $Z(\bPhi) $:}\\
    ~~$U \Sigma V^\dagger \gets \textrm{svd}(J^\intercal)$\\
    ~~$Z(\bPhi) \gets$ columns of $V$ corresponding to zero singular values\;

     \text{Find the kernel vector: }\\    
    ~~$\boldsymbol{\Delta} \bx  = -\nabla_\bx \mathcal{Q}( \bPhi + Z(\bPhi) \bx)$\\
    \text{or}\\
    ~~$\boldsymbol{\Delta} \bx = -Z(\bPhi)^T \nabla \mathcal{Q}(\bPhi)$.\\
    \text{Update parameters $\bPhi$: } \\
    ~~$\bPhi \gets \bPhi + s \frac{Z(\bPhi) \boldsymbol{\Delta} \bx}{\Vert Z(\bPhi)  \boldsymbol{\Delta} \bx\Vert }$ \\
    \uIf{$\mathcal{Q}(\bPhi) \leq \mathcal{Q}_\mathrm{aim}$}{
    break
    }
}
\end{algorithm}

\section{Numerical examples}
In this section, we demonstrate GECKO on several pulse-improvement tasks by choosing different quality functions $\mathcal{Q}(\boldsymbol{\Phi})$. We provide code to reproduce these examples at Ref.~\cite{our_data}.

In all our numerical experiments, we consider the following two-qubit Ising Hamiltonian with local transverse fields:
\begin{align}
    \label{eq:tfim_1}
    H(t) = g \sigma^z_1 \sigma^z_2 + h_1(t) \sigma^x_1 + h_2(t) \sigma^x_2,
\end{align}
where the local field strengths $h_1(t)$ and $h_2(t)$ are expressed in units of the coupling strength $g$. Unless stated otherwise, we set $g\Delta t=1$. We plot the time $t$ as the dimensionless ratio $t/T$ (except for Section~\ref{sec:pulse_duration_drift_constrained}).
Correspondingly, pulse amplitudes are plotted after multiplication by $T$, making them dimensionless. 
Different choices of $g\Delta t$ simply rescale the plotted amplitudes and frequencies. The Hamiltonian of Eq.~\eqref{eq:tfim_1} can be realized in ion traps~\cite{monroeProgrammableQuantumSimulations2021} and Rydberg atom arrays~\cite{morgadoQuantumSimulationComputing2021}. 
We also consider a second system in which the two qubits are driven simultaneously and qubit 2 has a constant local $z$-offset:
\begin{align}\label{eq:tfim_2}
    H(t) = g\sigma_1^z \sigma^z_2 + h_x(t)(\sigma^x_1 + \sigma^x_2) + h_z(t)(\sigma^z_1 + \sigma^z_2)  + \frac{1}{2}\sigma^z_2.
\end{align}
We consider two target unitaries. The first is the controlled-Z  gate:
\begin{align*}
    \operatorname{CZ} = \mathrm{diag}(1, i,i,1).
\end{align*}
The second is the CNOT gate:
\begin{align*}
    \operatorname{CNOT} =\begin{pmatrix}
        1 & 0 & 0 & 0 \\
        0 & 1 & 0 & 0 \\
        0 & 0 & 0 & 1 \\
        0 & 0 & 1 & 0 
    \end{pmatrix}.
\end{align*}
\subsection{\label{sec:spectral_filtering}Frequency-domain filtering}

Limited bandwidth is a common experimental constraint in quantum optimal control. Spectral constraints can be enforced directly during optimisation~\cite{lapert2009monotonically, palao2013steering, reich2014optimal}, incorporated into the pulse parameterisation, as in CRAB~\cite{canevaChoppedRandombasisQuantum2011}, imposed in GRAPE-based approaches~\cite{skinner2004reducing, kobzar2008exploring, li2011optimal}, or handled using automatic differentiation~\cite{song2022optimizing}.

Here, we show that frequency constraints can be added as a post-processing method using GECKO after a high-fidelity pulse has been found. 
To achieve this, we transform the pulse to the frequency domain and use a filter $H$ that multiplies each Fourier component by a weight $0\leq w\leq 1$.
The quality function $\mathcal{Q}_{\mathrm{filter}}(\bPhi)$ is then simply the distance between the original frequencies and the filtered frequencies:
\begin{align}
    \mathcal{Q}_{\mathrm{filter}}(\bPhi) =
    \left\|
    \mathcal{F}(\bPhi) - H(\mathcal{F}(\bPhi))
    \right\|_2^2.
\end{align}
In practice, we use a type-I discrete sine transform (DST-I), which preserves the boundary conditions of the original signal; see App.~\ref{app:dst1}.

\begin{figure}[htb!]
    \centering
    \includegraphics[width=\linewidth]{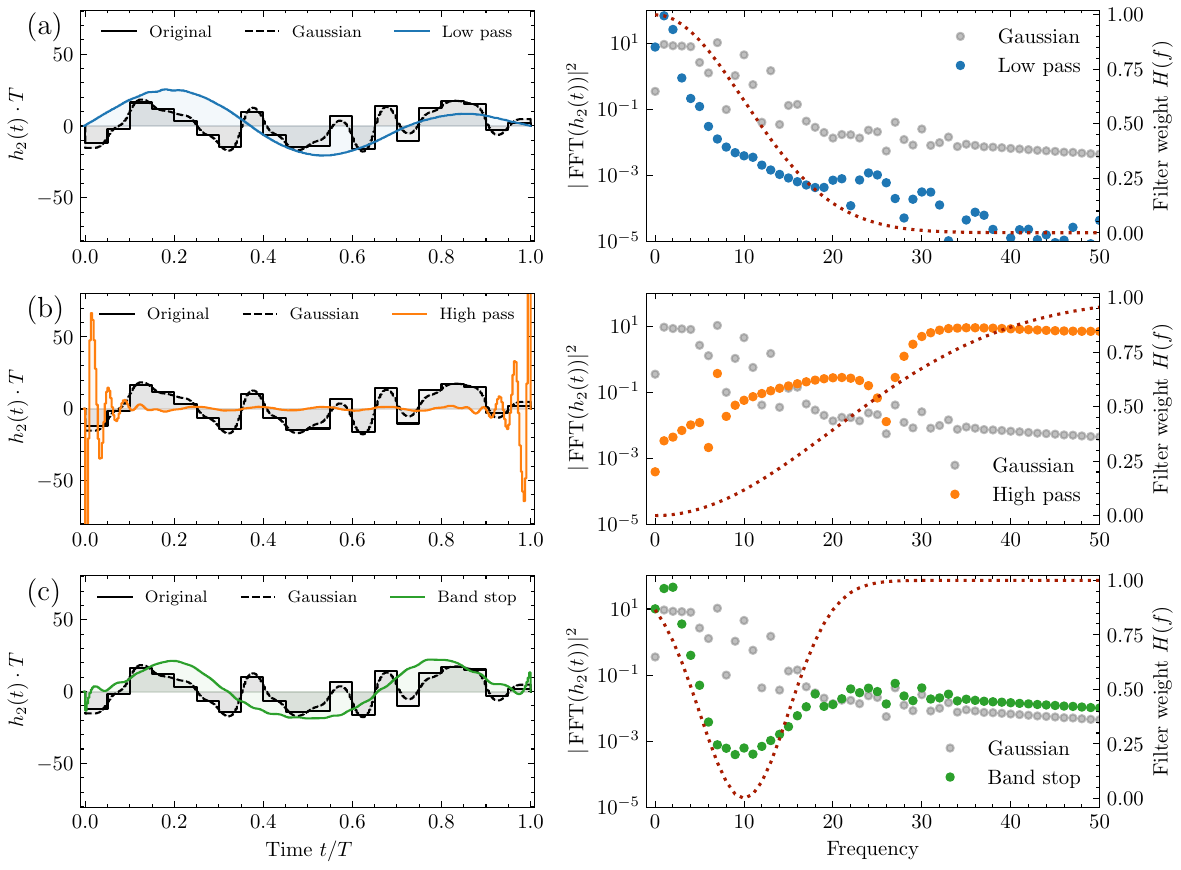}
    \caption{GECKO frequency filtering. The left column shows the original signal for $h_2(t)$ with $L=20$ as a black solid line. The dashed and coloured lines show the signal smoothed with a Gaussian filter of width $\sigma=4$ and the signal obtained with GECKO-based spectral filtering, respectively.
    The filtered signals are represented with $L=320$ piecewise steps. On the right, we show the power of the Fourier-transformed signal with the frequencies per unit time. The right y-axis indicates the filter weight $w\in[0,1]$. The grey dots indicate the power spectrum before filtering, and the coloured dots indicate the power spectrum after filtering. All signals reported here have fidelity $F\simeq0.999999$. (a) The low-pass filter is able to significantly reduce the high frequency components. (b) Conversely, the high-pass filter amplifies the higher frequencies while suppressing the low frequencies. (c) The band-stop filter suppresses frequencies around the centre of $f=10 g$.}
    \label{fig:filters}
\end{figure}

Note that the objective $\mathcal{Q}_{\mathrm{filter}}$ is differentiable because the transform and filtering operations are linear. We can thus obtain the gradient $\nabla \mathcal{Q}_{\mathrm{filter}}(\bPhi)$ and project it onto the kernel of the Jacobian using Eq.~\eqref{eq:optx}. In Fig.~\ref{fig:filters}, we show three examples of how 
GECKO can suppress or enhance selected frequency components while maintaining fidelity $F(\boldsymbol{\Phi}, U_{\mathrm{target}}) > 1 - \varepsilon$. We consider the Hamiltonian of Eq.~\eqref{eq:tfim_1} with $h_1(t)=0$, leaving a single control field acting on qubit 2.

\subsection{\label{sec:smoothing}Pulse smoothing}
Quantum optimal control methods often use piecewise-constant step functions for the control pulses. Such pulses can contain abrupt changes in the control parameters, which can increase the susceptibility to noise~\cite{motzoi_simple_2009, gambetta_analytic_2011}. The less smooth the pulse shape, the higher the frequency bandwidth required to generate the pulse, which poses an experimental challenge. To mitigate this, pulses are often smoothed either during the optimal-control search or as a post-processing step.
Frequency filter functions can be applied to smooth pulses of piecewise-constant steps~\cite{motzoiOptimalControlMethods2011, kirchhoffOptimizedCrossresonanceGate2018}, including Gaussian filters, which we use as a comparison to GECKO in the following. 

Intuitively, the smoothness of a pulse can be defined as the sum of squared differences between neighbouring time slices in a piecewise-constant pulse. We can construct a quality function that captures this notion of smoothness by introducing a linear operator $D$ that penalises changes between neighbouring time slices and returns the smoothing quality:
\begin{align}
    \mathcal{Q}_{\mathrm{smooth}}(\bPhi)
    &= \big\| D\,\bPhi \big\|_2^2, \qquad  D\,\bPhi = 
    \begin{pmatrix}
    \bphi_{1} - \bphi_{0}\\
    \bphi_{2} - \bphi_{1} \\
    \vdots \\
    \bphi_{L+1} - \bphi_{L} \\
    \end{pmatrix},\qquad \bphi_0 = \bphi_{L+1} = \boldsymbol{0},
\end{align}
which also enforces the experimentally desirable boundary condition $\bphi_0 = \bphi_{L+1} = \boldsymbol{0}$.
Substituting this objective into the direct kernel parameterisation of Eq.~\eqref{eq:Qminx} gives
\begin{align*}
     \min_{\boldsymbol{\bx}} \quad \mathcal{Q}( \bPhi + Z(\bPhi) \bx) =  \min_{\boldsymbol{\bx}} \quad \big\| D( \bPhi + Z(\bPhi) \bx) \big\|_2^2,
\end{align*}
This is a quadratic problem whose solution we can find directly with linear least-squares.

\begin{figure}[htb!]
    \centering
    \includegraphics[width=1\linewidth]{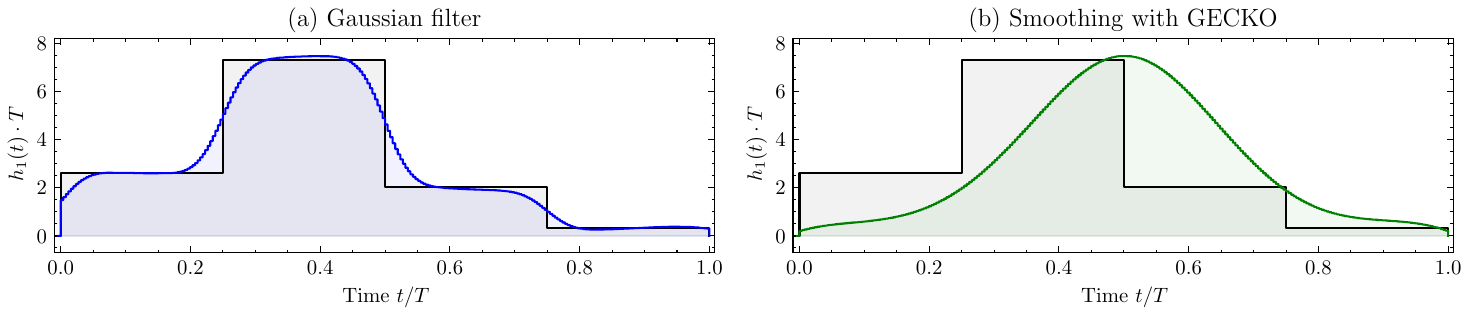}
    \caption{Control pulses for the local field $h_1(t)$ implementing a two-qubit CZ gate under the Hamiltonian in Eq.~\eqref{eq:tfim_1} with $h_2(t)=0$. The black line is the solution (infidelity $< \varepsilon = 1 \times 10^{-7}$) with 4 piecewise steps. The pulse is then smoothed (a) with a Gaussian filter (in the left plot in blue) and (b) with GECKO (in the right plot in green) and then optimised again to maintain a solution with infidelity $< \varepsilon$. The re-optimised solutions were found using GEOPE, but GRAPE or another quantum-control method could also be used.}
    \label{fig:gaussian_filter_vs_geometric_smoothing}
\end{figure}

In the first example, we set $h_2(t) =0$, and consider a single time-dependent control parameter $h_1(t)$ to implement the $\mathrm{CZ}$ gate. We find a solution with 4 piecewise steps and then smooth the pulse using either a Gaussian filter or GECKO, see Fig.~\ref{fig:gaussian_filter_vs_geometric_smoothing}. For the Gaussian filter, we subdivide each original time slice into 64 piecewise steps. We then pad the start and the end of the pulse with additional zeros. The Gaussian filter is applied with Gaussian kernel $\sigma=8$. After smoothing, we re-optimise to find a high-fidelity solution.
For GECKO smoothing, we iteratively double the number of time slices, apply a GECKO smoothing step, and then re-optimise the new parameters to maintain high fidelity. We perform this 6 times for a total factor of 64 additional piecewise steps, matching the discretisation used for the Gaussian-filter baseline. Note that GECKO produces a significant change in the pulse shape, approaching a very low-roughness solution with few extrema.
\begin{figure}[th!]
    \centering
    {\includegraphics[scale=0.62]{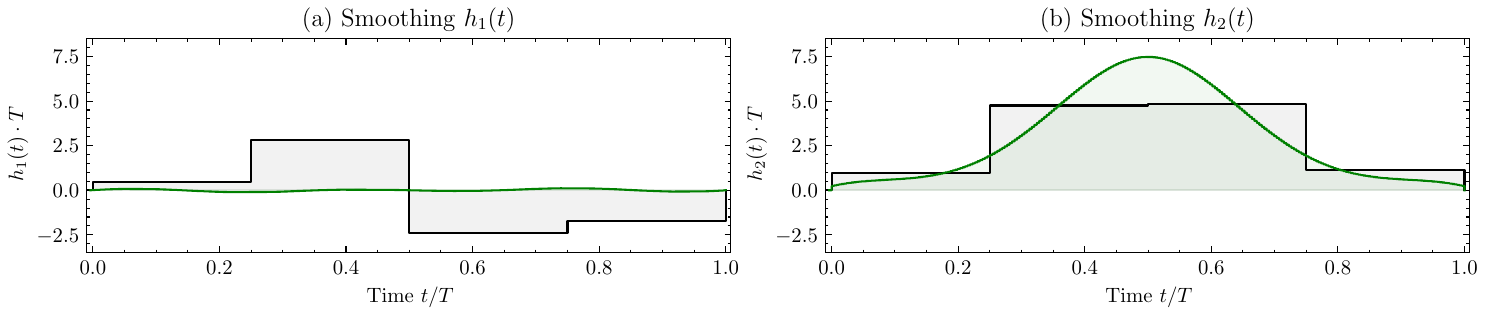}}
        
    \caption{Control pulses for the local fields (a) $h_1(t)$ and (b) $h_2(t)$ implementing a two-qubit CZ gate under the Hamiltonian in Eq.~\eqref{eq:tfim_1}. The initial solution with 4 piecewise steps is given in black and the solution given by smoothing with GECKO is given in green. The solution maintains an infidelity $\varepsilon<1\times10^{-7}$. 
    After smoothing, GECKO effectively eliminates the need for the $h_1(t)$ control.}
    \label{fig:smoothed_example}
\end{figure}

GECKO smooths the pulse by moving globally within the fidelity-preserving directions, rather than merely rounding local discontinuities. As a result, unnecessary control components can be suppressed when doing so lowers the global roughness objective $\mathcal{Q}_{\mathrm{smooth}}(\bPhi)$. As an example of this, we now allow for a non-zero $h_2(t)$ control parameter. In practice, we see that GECKO often finds the smoothed solution where one of the control parameters is not required, as demonstrated in Fig.~\ref{fig:smoothed_example}, because the global smoothness is maximised. A Gaussian filter on the other hand, will never find this solution.

\begin{figure}
    \centering
    \includegraphics[width=\linewidth]{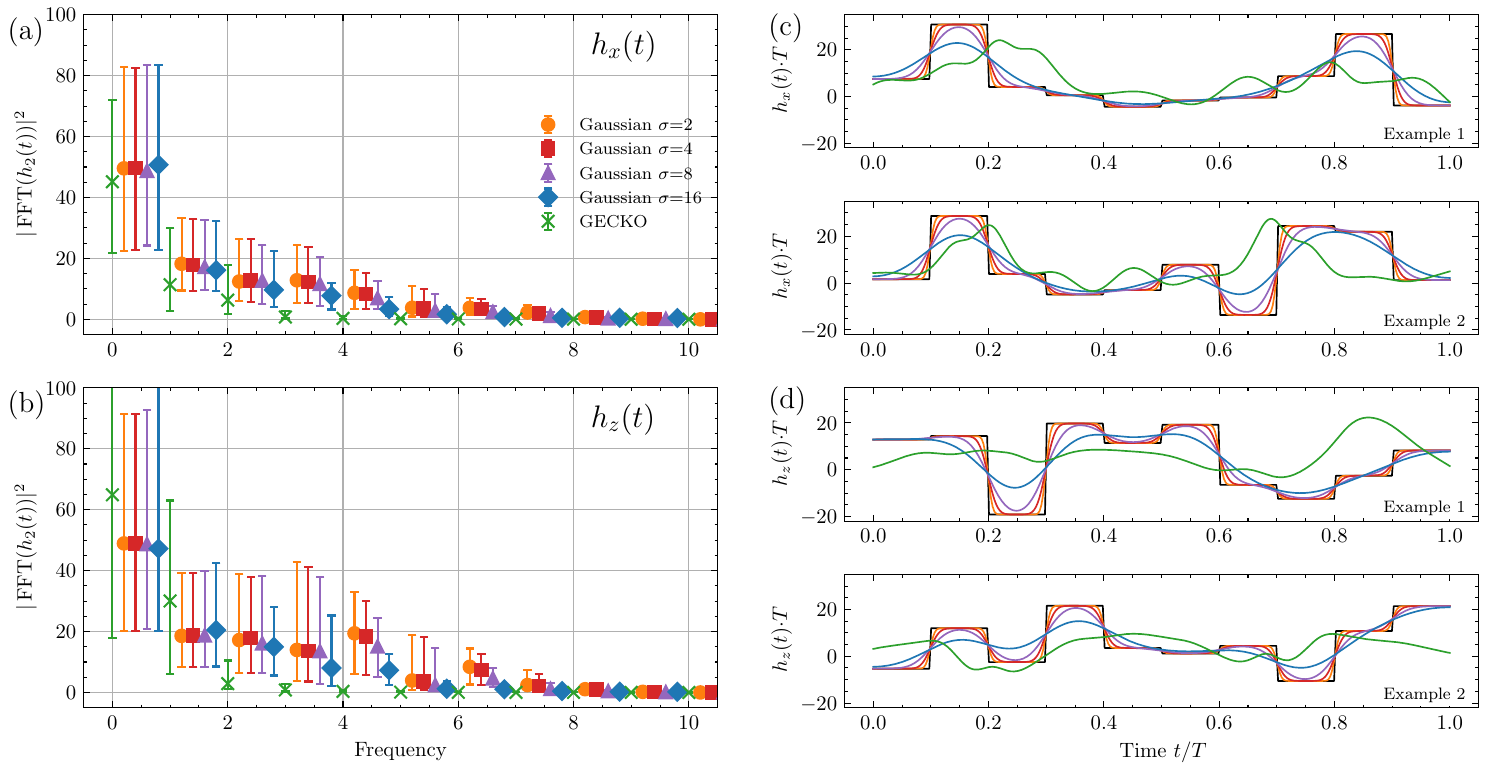}
    \caption{GECKO pulse smoothing over 100 independently initialised solutions with $L=10$ piecewise steps. Each solution is refined to $L=320$ and smoothed either with Gaussian filters of different widths $\sigma$ or with GECKO. The two panels on the left show the median power of the resulting pulses $h_x(t)$ and $h_z(t)$ averaged over the 100 initialisations. The error bars indicate the 25th and 75th percentiles. The two panels on the right show two examples of the resulting pulses found by the different smoothing procedures. All pulses shown have an infidelity of $\varepsilon<1\times10^{-4}$.}
    \label{fig:spectrum}
\end{figure}

As discussed above, high-frequency components can increase noise sensitivity and are difficult to implement under bandwidth limitations. 
An advantage of GECKO over Gaussian filtering is that optimising the global smoothness objective tends to reduce high-frequency spectral components more effectively.
This is shown in Fig.~\ref{fig:spectrum}, where we consider the CNOT as target gate with the Hamiltonian Eq.~\eqref{eq:tfim_2}. 

\subsection{\label{sec:robust_pulse}Noise-robust pulses}
A desirable control-pulse property is robustness of the fidelity to small parameter deviations. Previously, filter functions in GRAPE have been used for parameter insensitivity~\cite{jiaAnglerobustTwoqubitGates2023, kangDesigningFilterFunctions2023}, as well as geometric methods~\cite{barnes2022dynamically,albertini2026quantum}. We want the parameters corresponding to the set of $K' \leq K$ generators $\mathcal{H}' \subseteq \mathcal{H}$ to be robust to fidelity deviations. 
To evaluate robustness, we compute the fidelity after adding a perturbation $\Delta_{l,k}$ to each affected control amplitude. The shift is the same for every piecewise step, so $\Delta_{1,k} = \Delta_{2,k} = \cdots = \Delta_{L,k} = \Delta_{k}$. 
The perturbation vector $\bm{\Delta}$ is therefore defined componentwise by
\begin{align}
    \Delta_{l,k} = \begin{cases}
        \Delta_k & \textrm{ if } G_k \in \mathcal{H}', \\
        0 & \textrm{ otherwise. }
    \end{cases}
\end{align}
Each $\Delta_k$ lies in a range defined by a hyperparameter $\delta$, giving $\Delta_k \in [-\delta, \delta]$ for each $k$ index corresponding to $G_k \in \mathcal{H}'$. We define the robustness objective as the worst-case fidelity over this perturbation set:
\begin{align}
    \mathcal{Q}_\textrm{robust}(\bPhi) = 1 -  \min_{\bm{\Delta}}\:  F(\bPhi + \bm{\Delta} , U_\textrm{target}) \textrm{ where } \Delta_k \in [-\delta, \delta]~\textrm{for all}~k~\textrm{s.t.}~G_k \in \mathcal{H}'. 
\end{align}
To make this problem computationally tractable, we select $S$ points, uniformly spaced, in the range $[-\delta,\delta]$ for every robust parameter, giving a total of $S^{K'}$ points. 

\begin{figure}
    \centering
    \subfloat[Fidelity]{\includegraphics[width=0.6125\linewidth]{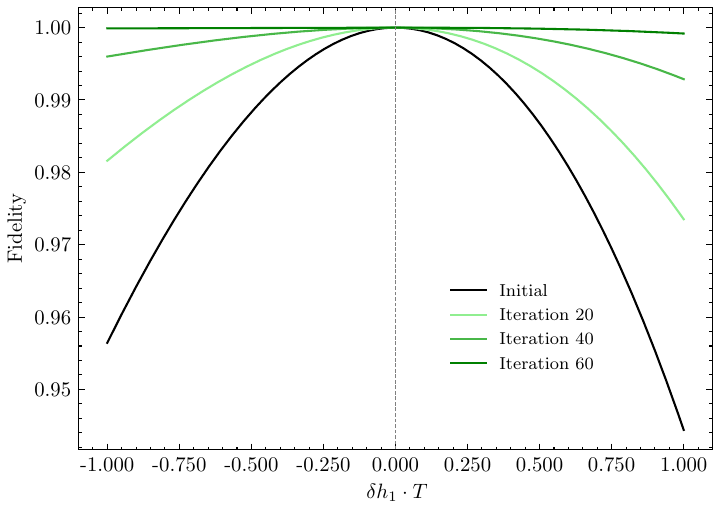}}
    \hspace{0.02\linewidth}
    \subfloat[Parameters]
    {\includegraphics[width=0.358\linewidth]{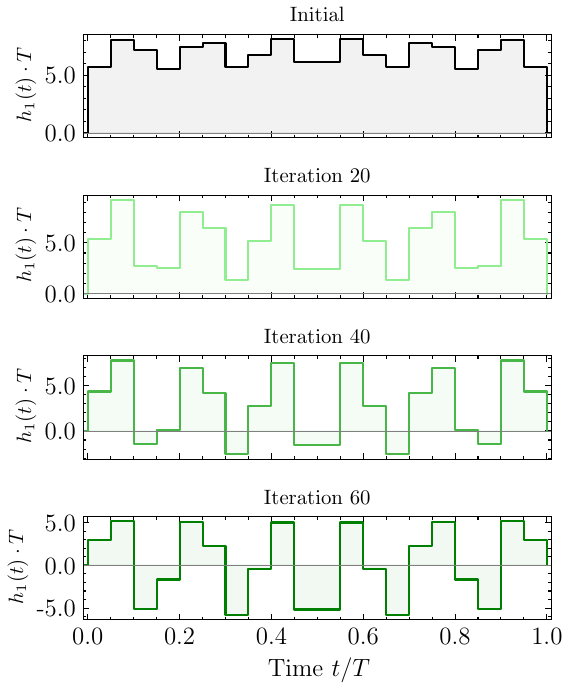}} 
    \caption{The robustness quality function is optimised with GECKO for a CZ gate pulse with the model of Eq.~\eqref{eq:tfim_1} where $h_2(t) = 0$ and $L=20$ piecewise steps. (a) Fidelity is plotted against the small deviation $\delta h_1$ of the solution parameter $h_1(t)$. After 20 iterations of GECKO, the pulse is re-optimised with GEOPE to ensure a high fidelity solution. (b) The parameters after each of the 20 iterations of GECKO are shown.}
    \label{fig:robustness}
\end{figure}

As a numerical example, we return to the transverse-field Ising model of Eq.~\eqref{eq:tfim_1} used in Section~\ref{sec:smoothing}. The results are shown in Fig.~\ref{fig:robustness}. 
We set $h_2(t)=0$, leaving a single control field $h_1(t)$ to implement a two-qubit CZ gate. We choose $L=20$ piecewise steps to give enough freedom for finding robust pulses. 
We set the perturbation scale to $\delta=0.05$, corresponding to $\delta T=1.0$ for this example, which is large enough to noticeably reduce the fidelity of the initial GEOPE solutions. Having found an initial solution with GEOPE, we use GECKO with our robustness quality function $\mathcal{Q}_{\mathrm{robust}}(\bPhi)$. Every 20 iterations of GECKO, we use GEOPE again to increase the fidelity. After a total of just 60 iterations of GECKO, we find a significantly more robust pulse.

\subsection{\label{sec:time_optimsation}Pulse duration}

Short pulse duration is an important criterion in quantum optimal control~\cite{ndong2014time, janduraTimeOptimalTwoThreeQubit2022}. A faster gate can be beneficial for two important reasons. Firstly, it allows more gates to be performed and therefore more computation per unit time. Secondly, qubits have a finite coherence time, hence slow gates can increase the probability of errors.

In the absence of platform-specific constraints, one can view the minimum evolution time geometrically as a shortest-path problem on the manifold $\SU{N}$, in analogy with complexity geometry~\cite{nielsenGeometricApproachQuantum2005, brownQuantumComplexityLower2023, brownPolynomialEquivalenceComplexity2024}.
If the available Hamiltonian norm is fixed, the path length becomes equivalent to the pulse duration. We first consider this case, where we minimise the geometric path length.

In practice, the entangling terms determine the minimum time for the pulse. For the Hamiltonians we consider here, the entangling term is a single drift term $g\sigma_1^z\sigma_2^z$. Previously, we set $g \Delta t = 1$. In this section, we set the timescale of the evolution by fixing the coupling term $g$, such that the time is given in units $1/g$. The total pulse duration is $T = L \Delta t$ and we now have an additional pulse parameter, $\Delta t$. We consider both the geometric and drift-term time optimality.

\subsubsection{\label{sec:pulse_duration_path_length}Minimum path length}
\begin{figure}
    \centering
    \includegraphics[width=\linewidth]{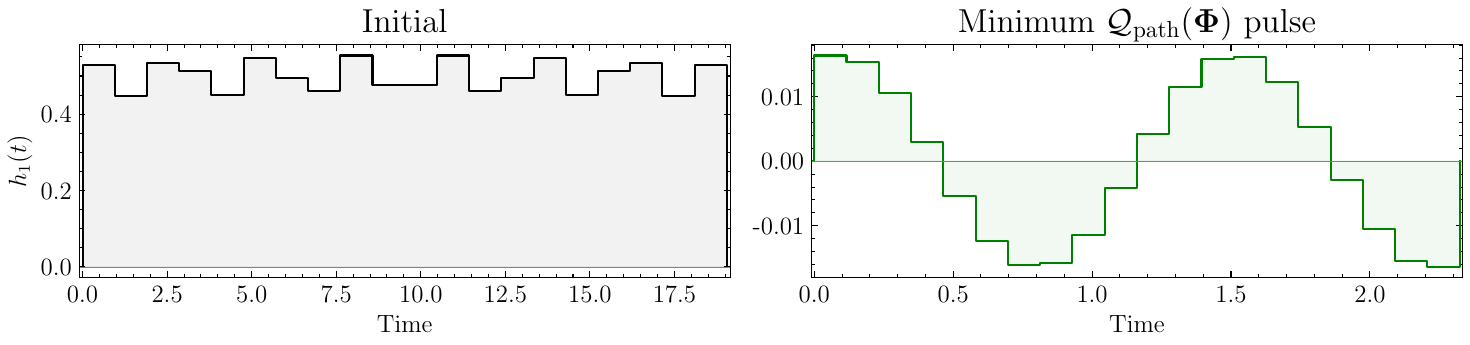}
    \caption{CZ pulse under the Hamiltonian in Eq.~\eqref{eq:tfim_1} with $h_2(t)=0$ and $L=20$ piecewise steps. (a) The initial pulse after optimal control (GEOPE). (b) The final pulse after GECKO is used to minimise the pulse duration function $\mathcal{Q}_\textrm{path}(\bPhi)$. The total time is reported in units of the fixed drift coupling strength $g$, giving $T=2.323/g$.}
    \label{fig:init_vs_min_duration}
\end{figure}

The minimum path length is the case for which there is limited experimental capability for the Hamiltonian control terms. Thus, the total Hamiltonian norm must be minimised. There is an equivalence between path length on the $\SU{N}$ manifold and the duration of the pulse~\cite{lewis2025quantum}. A time-independent Hamiltonian gives a path length on the manifold that can be found straightforwardly~\cite{wiersemaHereComesSUN2024}. Defining our metric $g(x,y)=\Tr{x^\dagger y}/N$, the length of the path of each $l$th piecewise step is
\begin{align}
    L[U(\bphi_l)] = \sqrt{\bphi_l \cdot \bphi_l + g^2} \Delta t, 
\end{align}
where the parameters $\bphi_l$ include the control terms and $g$ is the strength of the drift term. We then take the total path length as the pulse quality function
\begin{align}
    \mathcal{Q}_\textrm{path}(\bPhi) = \sum_{l=1}^{L}\sqrt{\bphi_l \cdot \bphi_l + g^2} \Delta t,
\end{align}
which we minimise at fixed drift strength $g$ while allowing $\Delta t$ to vary.  
We use GECKO to optimise the CZ gate using the model of Eq.~\eqref{eq:tfim_1} with $h_2(t)=0$. Across many random initialisations, GECKO converges to the same minimum path length; one representative example is shown in Fig.~\ref{fig:init_vs_min_duration}. The resulting pulse gives a total duration $T=2.323/g$ for this case. 

\subsubsection{\label{sec:pulse_duration_drift_constrained}Timescale determined by drift term}
The timescale can be determined by the strength of a single Hamiltonian term: the strength of the constant drift.
Experimentally, this interacting drift term often cannot be changed, and determines the timescale of the pulse. 

\begin{figure}
    \centering
    \includegraphics[width=\linewidth]{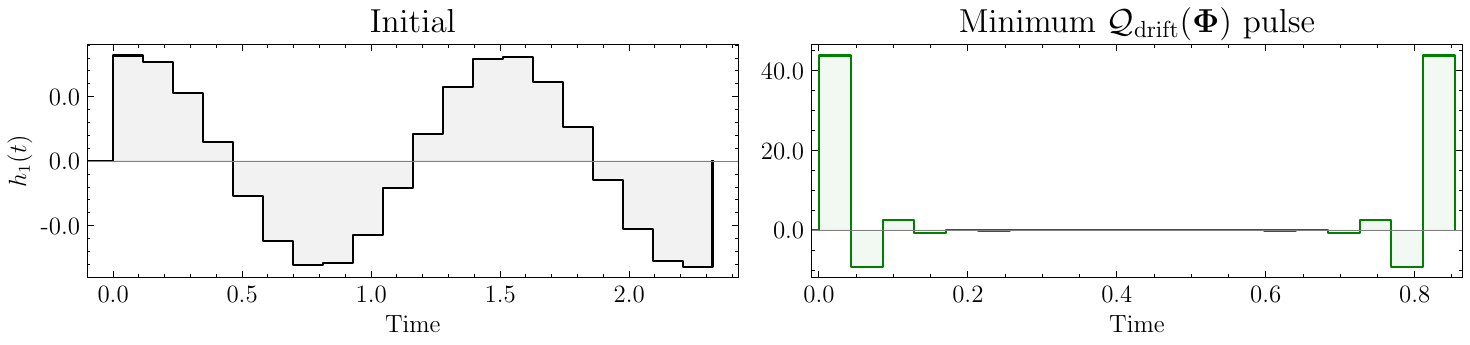}
    \caption{The pulse solution for a CZ gate pulse with the model of Eq.~\eqref{eq:tfim_1} where $h_2(t) = 0$ and $L=20$ piecewise steps. (a) The initial pulse is the GECKO path-length-optimised solution shown in Fig.~\ref{fig:init_vs_min_duration}. (b) The final pulse after GECKO minimisation of $\mathcal{Q}_{\mathrm{drift}}$. The total time is reported in units of the fixed drift coupling strength $g$, giving $T=0.854/g$. }
    \label{fig:init_vs_min_drift}
\end{figure}

In this case, $g$ is fixed to determine the timescale. We then minimise the $\Delta t$ and the other control parameters $\bPhi$ can become arbitrarily large. The pulse quality function is simply  
\begin{align}
    \mathcal{Q}_\textrm{drift}(\bPhi) = \Delta t,
\end{align}
which, when minimised, gives the minimum pulse duration. We optimise $\mathcal{Q}_\textrm{drift}(\bPhi)$ for a CZ gate, using, as previously, the model of Eq.~\eqref{eq:tfim_1} with $h_2(t)=0$. We converge on a minimum for $\mathcal{Q}_\textrm{drift}(\bPhi)$ by initialising with the solution of the path length, $\textrm{argmin}_{\bPhi} \left( \mathcal{Q}_{\textrm{path}}(\bPhi) \right) $, an example is shown in Fig.~\ref{fig:init_vs_min_drift}. The total pulse duration is $T=0.854/g$. While this time is considerably shorter than minimising the path length, the maximum local term $h_1(t)$ is about $3\times 10^3$ larger than the path-length optimised controls of $\mathcal{Q}_\textrm{path}(\bPhi)$. Note that we do not have to initialise $\bPhi$ as the solution of the path length case. However, if we take a random initialisation, the minimisation for $\mathcal{Q}_\textrm{drift}(\bPhi)$ can get trapped in suboptimal local minima.

\section{\label{sec:discussion}Summary and Conclusion}
We have introduced geometric quantum control with kernel optimisation (GECKO), a method for improving quantum control pulses while preserving the implemented unitary to first order. GECKO exploits the Riemannian geometry of the special unitary group $\mathrm{SU}(2^n)$ for an $n$-qubit system. Starting from a high-fidelity pulse, the method identifies directions in pulse-parameter space that lie in the kernel of the parameter Jacobian. Updates along these directions leave the implemented unitary unchanged to first order,
and can therefore be used to optimise secondary pulse properties while maintaining high fidelity.

A key advantage of GECKO is that it separates the task of finding a high-fidelity pulse from the task of improving its experimental properties. 
Rather than imposing all hardware-motivated constraints during the initial control search, one can first find any suitable solution and then move along the corresponding fidelity level set to optimise a chosen pulse-quality objective. This objective can be any differentiable function of the pulse parameters. 
We demonstrated this flexibility using objectives for spectral filtering, pulse smoothing, robustness to parameter deviations, and pulse-duration reduction. 
Code implementing these examples is available at Ref.~\cite{our_data}.

It is possible, and in many cases beneficial, to optimise over a number of pulse features simultaneously. This can be achieved by either performing GECKO for the various quality functions sequentially or by using a pulse quality function that is a weighted sum of quality functions, $\mathcal{Q}(\bPhi) = \sum_i w_i\mathcal{Q}_i(\bPhi)$. The weights $w_i$ determine the relative importance of optimising each feature of the pulse. Future work can investigate this idea in relation to experimental pulse design for more realistic systems than we considered here. 

In our examples, it appears quite possible that the pulse found is a global minimum of the quality function. This is particularly likely in the case of the smoothing example in Section~\ref{sec:smoothing} because the smoothest pulse is likely to have only one extremum. In the case of minimising pulse duration by minimising the path length on the manifold, GECKO can be a method for solving the geodesic equation in the subRiemannian manifold, which corresponds to a restricted geometry where only certain tangent space directions are available~\cite{agrachev2019comprehensive}. If GECKO converges to a solution, we have reached an extremum of the path length functional for a fixed number of piecewise steps. Repeated uses of this method with increasing numbers of piecewise steps could converge on a solution to the length of the geodesic. This has implications in complexity geometry, where the minimum path length is equivalent to the minimum circuit complexity~\cite{nielsenGeometricApproachQuantum2005, nielsenQuantumComputationGeometry2006, dowlingGeometryQuantumComputation2008, brownQuantumComplexityLower2023, brownPolynomialEquivalenceComplexity2024}. 

We have shown the efficacy of GECKO for finding optimal pulses for 2-qubit gates. 
Although the formulation applies to any number of qubits, its current implementation is limited by the cost of computing the Jacobian and its kernel, which grows rapidly with the number of qubits. This is not prohibitive for a small number of qubits ($\lesssim 5$ qubits), but, as with quantum optimal control methods broadly, it cannot be expected that a pulse for a larger number of qubits ($\gtrsim 5$ qubits) can be optimised straightforwardly. Improving the efficiency of each GECKO step by exploiting symmetries or the sparsity of the Jacobian is left for future work. The trade-off between a larger step size $s$ giving faster convergence, and the second-order error that accumulates during GECKO is also of practical importance. The optimal choice of step size and re-optimisation schedule will likely depend on the details of the experimental platform.

\section{\label{sec:acknowledgements}Acknowledgements}

The authors contributed equally. The Flatiron Institute is a division of the Simons Foundation.

\clearpage
\appendix

\section{Discrete Sine Transform of type I}\label{app:dst1}

The discrete sine transform of type I (DST-I) is implemented by embedding the signal into an odd extension and then applying a real Fourier transform. For a fixed channel \(k\), let
\begin{align*}
    \bphi^{(k)}
    :=
    (\phi_{1,k},\dots,\phi_{L,k})\in\mathbb{R}^{L}.
\end{align*}
We construct the odd extension
\begin{align*}
    \boldsymbol{\varphi}^{(k)}
    :=
    \bigl(0,\phi_{1,k},\dots,\phi_{L,k},0,-\phi_{L,k},\dots,-\phi_{1,k}\bigr),
\end{align*}
so that $\boldsymbol{\varphi}^{(k)}$ has length $2(L+1)$. This extension is antisymmetric about both endpoints, which enforces vanishing boundary values.

The DST-I coefficients are then obtained from the imaginary part of the Fourier transform of $\boldsymbol{\varphi}^{(k)}$:
\begin{align*}
    \widehat{\boldsymbol{\varphi}}^{(k)}_n
    :=
    -\mathfrak{Im}\left[\operatorname{FFT}(\varphi^{(k)})_n\right],
    \qquad n=1,\dots,L.
\end{align*}
Here the mode $n=0$ is omitted, since it corresponds to the DC component. Equivalently, these are precisely the coefficients of the discrete sine-series expansion
\begin{align*}
    \phi_{t,l}
    =
    \sum_{n=1}^{L}
    \widehat{\varphi}_{n,k}\,
    \sin\!\left(\frac{\pi n t}{L+1}\right),
    \qquad l=1,\dots,L,
\end{align*}
with coefficients
\begin{align*}
    \widehat{\varphi}_{n,k}
    =
    \frac{2}{L+1}
    \sum_{l=1}^{L}
    \phi_{l,k}\,
    \sin\!\left(\frac{\pi n l}{L+1}\right),
    \qquad n=1,\dots,L.
\end{align*}
Thus, for each channel $k$, the signal is expanded in a basis whose modes vanish at the endpoints $l=0$ and $t=L+1$. Consequently, filtering in this representation preserves zero-boundary conditions.

\end{document}